\documentclass{aa} 
\usepackage{graphicx}
\usepackage{txfonts}
\bibpunct{(}{)}{;}{a}{}{,} 

\def\kms {km\,s$^{-1}$}

\begin{document}

   \title{Making sense of the spectral line profiles of Betelgeuse and other Red SuperGiants\thanks{Based on observations obtained at the T\'elescope Bernard Lyot
(TBL) at Observatoire du Pic du Midi, CNRS/INSU and Universit\'e de
Toulouse, France.}}

   \author{{ A.~L{\'o}pez Ariste}\inst{1}, { Q.~Pilate}\inst{1},{ A.~Lavail}\inst{1},{ Ph. Mathias}\inst{2},  }
   \date{Received ...; accepted ...}

   \institute{IRAP, Universit\'e de Toulouse, CNRS, CNES, UPS.  14, Av. E. Belin. 31400 Toulouse, France 
   \and IRAP, Universit\'e de Toulouse, CNRS, CNES, UPS. 57, Av. d'Azereix. 65000 Tarbes, France
}

 
  \abstract{
  
  Spectropolarimetry of atomic lines in the spectra of Betelgeuse, and other Red SuperGiants (RSG), presents broad  line profiles in 
  linear polarization, but narrow profiles in intensity. Recent observations of the Red SuperGiant RW Cep show, on the other 
  hand, broad intensity profiles, comparable to those in linear polarization. This observation hints that this  difference in the Stokes Q/U and I profile 
  widths noted in many RSG is just a temporary situation of the atmosphere of a given star. We propose an explanation 
  for both cases based on the presence of strong velocity gradients larger than the thermal broadening of the spectral line. Using 
  analytical but very simple radiative transfer we can compute intensity line profiles in such scenarios. We find that they qualitatively 
  match the observed broadenings: large gradients are required for the narrow profiles of Betelgeuse, small gradients for the broad profiles 
  of RW Cep. Profile bisectors are also reasonably well explained by this scenario in spite of the simple radiative transfer treatment. 
  
  These results give a comprehensive explanation of both intensity and polarisation profiles. They provide also comfort to the approximation
  of a single-scattering event used in explaining the observed linear polarization in the inference of images of the photosphere of Betelgeuse and 
  other Red SuperGiants.

  The atmospheres of Red SuperGiants appear to be able, perhaps cyclically, to either produce large velocity gradients that prevent photospheric plasma 
  from reaching the upper atmosphere and that must hinder large events of mass loss, either to  leave vertical movements unchanged, letting 
  plasma  raising, escaping gravity and forming large dust clouds in the circumstellar environement. 
  The origin of the velocity gradient and its modulation within the atmosphere remains an open question. }
  

\titlerunning{Making sense of RSG spectra}
\authorrunning{A. L\'opez Ariste et al.}

   \maketitle

\section{The issue with the spectra of Red SuperGiants.}
The discovery of linear polarisation in the atomic lines of the spectrum of Betelgeuse \citep{auriere_discovery_2016}  opened the path to a succesful imaging technique 
that has provided continuous images of the convective structures in the photosphere of Betelgeuse (since 2013) and other Red SuperGiants (RSG) \citep{lopez_ariste_convective_2018}. More recently, exploiting 
the different formation heights of different atomic lines, 3-dimensional images of the photosphere have been inferred from those spectropolarimetric 
signals \citep{lopez_ariste_three-dimensional_2022}. Beyond the images themselves, this technique has unveiled the large convective velocities predicted by numerical simulation codes and confirmed 
the size and temporal scales of variation of these convective structures. The measurement of plasma velocities at different heights has also uncovered 
plasma plumes that barely change velocity as they rise through the atmosphere of Betelgeuse, pointing to the presence of an acceleration mechanism 
hitherto unidentified but present from the photospheric heights and able to equilibrate forces with gravity and maintain plasma velocities as 
they approach the height at which they will definitely escape as stellar wind.

These successes were founded in a series of approximations about the polarized line formation which have been discussed and partially justified 
in the cited literature. Among them, perhaps the weakest is the imposition of a relationship between brightness and vertical velocity. Such correlation 
is observed whenever the line formation happens while convection is still the dominant dynamical process, as in the Sun. Numerical simulations of the 
photosphere of Betelgeuse \citep{kravchenko_tomography_2019} apparently infirm such relationship and suggest that the  formation region of most of the atomic lines present in 
the spectrum resides well above the convective zone, and brightness 
and velocity would not be clearly correlated in the formation of spectral lines in Betelgeuse. Against this prediction stands the comparison of 
images inferred from spectropolarimetry assuming this correlation true, and the reconstructed images obtained through interferometry. 

\cite{kravchenko_tomography_2018} explored whether convection still leaves a signature in the intensity spectra and, through a newly-developed tomographic technique, unveiled 
a histeresis loop reminiscent of what could be expected if lines were formed in convective conditions, that is when the correlation of brightness and 
vertical velocities is still valid. Such histeresis loops could be seen in observed spectra and also, marginally, in synthetic spectra computed 
by radiative transfer through stellar snapshots  numerically simulated. But these same authors claimed that such histeresis loops may be understood 
not as reflecting that lines form where convection takes place, but rather that acoustic waves or pulsations may propagate correlations of the 
kind to selected patches well above where convection imposes those relationships between brightness and vertical velocity. 
Since, traditionally, convection has also been called for to explain the C-shape line asymmetries \citep{gray_mass_2008}, it should go without 
further justification that 
it is in the intensity profile that we should seek the answer to the validity of the brightness-velocity correlation at the core of the interpretation
of the measured linear polarization in terms of images of convective structures. 

\begin{figure*}
   \includegraphics[width=\textwidth]{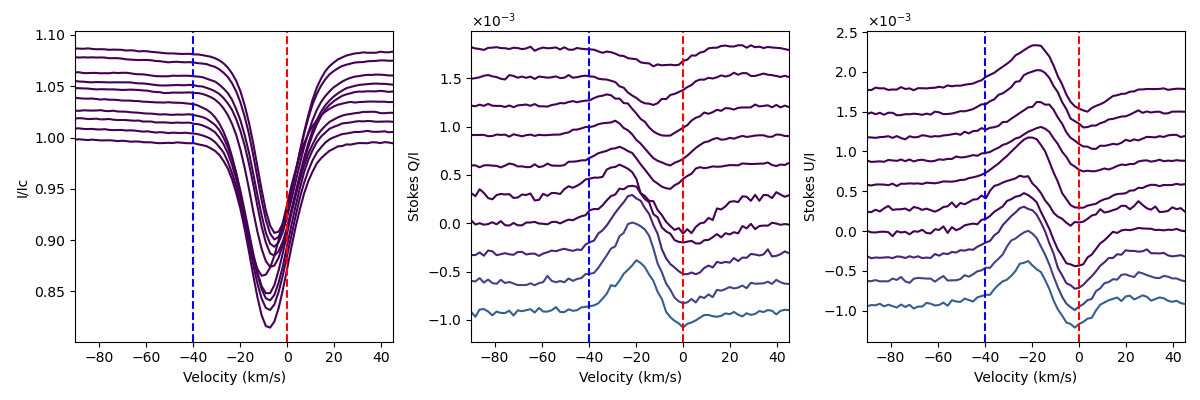}
   \caption{Observed profiles of Betelgeuse in the period 27 November 2013 through 3 March 2015  with Narval at the TBL. 
   The red vertical dashed line marks the chosen reference velocity $v_*$ interpreted as the velocity of the center of mass of the star. The blue 
   dashed line marks the maximum vertical velocity $v_{max}$ of the plasma. }
   \label{obsprofiles}
   \end{figure*}

However, inspection of the observed intensity and linear polarization profiles makes it clear that there is a further problem with intensity profiles 
than just correctly interpreting bisectors or hysteresis loops. As emphasized by \cite{lopez_ariste_three-dimensional_2022}, 
the intensity profiles observed on Betelgeuse are astonishingly 
narrower than the width of the linear polarization signal. This is illustrated in Fig. \ref{obsprofiles}, where the limits of $v_*$, the assumed velocity 
of the star, and $v_{max}$, the assumed highest velocity of the rising plasma are also shown. These two limits are set by simple inspection of the
accumulated data set of linear polarization profiles. In a model of pure convective movements, all that is bright is rising vertically, and the most 
blushifted signal observed must come from plasma rising close to disk center at the maximum velocity possible. Thus, no linear polarization is 
to be observed at shorter wavelengths than this $v_{max}$ limit. Similarly, plasma at the limb of the star emits at zero velocity respect to the 
bulk star's velocity $v_*$. This red limit is not strict: sinking plasma, though darker than rising plasma, produces signals redder than this limit. 
But these zones should lead to very low emission.
\cite{lopez_ariste_height_2023} found that the spectra of the RSG $\mu $ Cep presented strong polarization signals beyond the red wing of the line profile and 
interpreted them as due to the intermittent plumes arising in the back hemisphere of the star, but rising high enough to 
be visible above the limb. But beyond these episodic phenomena, no large signals are observed to the red. \cite{lopez_ariste_three-dimensional_2022} 
discussed at length the validity and justification of these two velocities  for Betelgeuse.  In what concerns here, nevertheless, it is not the actual 
value of those two velocities, but the  most obvious that whatever their precise value  there is a considerable amount of linear polarization 
arising at wavelengths which, judging from the intensity profile alone,  are in the continuum, beyond the blue 
wing of the line. It is difficult to make much sense of this. Continuum is precisely defined by its quasi-independence of wavelength, at least at the 
scales here concerned. The observed linear polarization signal varies rapidly with wavelength and cannot be due to any continuum emitting process. Interpreting it as the polarisation of the extreme 
wing of the spectral line makes no much sense either. It would require polarisation levels of 10\% at some wavelengths, which is also nonsense. There is no 
mechanism by which such high levels of linear polarization can be created in the extreme wing of the line, when elsewhere in the line the observed polarisation amplitudes 
are  0.1\%  at most. 

This is the problem we address in the present work. We propose a mechanism by which the intensity profile, after integration over the disk, is narrowed 
while the linear polarisation profile keeps the original width. Such mechanism is based upon two ingredients. The first one is  that in the absence of stellar rotation,
and in an atmosphere where plasma presents predominantly vertical velocities, integration
over the disk favours signals emerging from the disk center, since the wavelength of emission is modulated by its projection onto the line of sight, and the
$\cos \theta$ function reaches a maximum at disk center. Hence velocities around  disk center, if relatively homogeneous, are similar and add up together, while at 
the limb a small change in position considerably changes the projected velocity, even if the atmosphere is homogeneous. This effect is common to all disk integrations, so, alone, 
it makes no difference between Betelgeuse and any other star. The second ingredient is critical: it requires the presence of gradients of velocity 
in the region of formation of the line larger than the width of the local line. These gradients deformate the line profiles giving them a triangular 
shape \citep{bertout_line_1987}. Due to it, and in a nutshell: regions at disk center with large gradients produce  triangular shape profiles that contribute to all wavelengths, while 
regions at the limb with low gradients produce narrow gaussian profiles centered at near zero velocities, around $v_*$. The correct addition of these contributions 
produces the observed profiles as we shall demonstrate in the next section. 

In the process to understand the intensity profiles of Betelgeuse, a key element has been the observation of another star, RW Cep. This star 
is not particularly different from Betelgeuse. Perhaps it is better identified as a yellow hypergiant \citep{anugu_time_2024}. But the characteristics and evolution 
of the photosphere are identical and any observational differences between Betelgeuse and RW Cep, or any other similar evolved giant star,
must be mostly attributed to whatever the star is doing right now: is it quietly convecting as Betelgeuse appears to do (except perhaps during 
the recent great dimming event), or is on the contrary RW Cep in a Decin stage as, for example, $\mu$ Cep, with repeated events of large plumes raising sufficiently high to escape 
gravity and form a new shell of dust clumps \citep{lopez_ariste_height_2023,decin_probing_2006}. It was a big surprise to discover therefore 
that RW Cep presents in our 2023 observations a large and 
broad intensity profile, fully compatible with the linear polarisation signals and unlike anything observed in Betelgeuse in the last 10 years. 
RW Cep is  doing something right now which broadens the profiles. We were forced to reach the conclusion that atmospheric dynamics, in its 
broadest sense, is  able to either broaden or narrow intensity 
profiles. The narrow profiles observed in Betelgeuse are not intrinsic to an RSG, they are rather due to the present dynamics of the star. 
This realisation opened the path 
to our proposed solution. In Section 3 we explore how our proposition based upon velocity gradients can explain both Betelgeuse and RW Cep intensity 
profiles.

Two last observational facts are discussed and modelled in Section 4. Bisectors of Betelgeuse line profiles often present a typical C-shape, but 
from time to time they reverse their asymmetry. 
Our proposed solution must explain this difference and their change in time. And finally, the intensity line presents a velocity span of 7-10\,\kms 
over time. Since linear polarisation signals confirm the 40\,\kms velocities  predicted by numerical simulations to be present in the convective atmosphere, 
this  velocity span cannot be a direct projection of the dynamics of the atmosphere but rather the result of some kind of averaging that our solution 
must explain as well.

\section{Radiative transfer in the photosphere of an RSG.}

Our approach to the problem of radiative transfer in these atmospheres responds to a constraint and a desire. The constraint is that, in spite 
of all its successes and of the harbinger of ever greater and precise details, we cannot consider numerical simulations as the ultimate description 
of red supergiants. Not just that sufficient simulations with the appropriate parameters are not yet available but also that it is our purpose to always 
check those models and not to assume them true and limit observations to a perennial validation of the models. The desire is for an explanation rather 
than a description or a perfect fit. We do not desire at this point to reproduce the intensity profiles quantitatively, but to identify the key physical 
ingredients responsible of the main observed features. 

With those criteria in mind, we can boldly split the problem of radiative transfer in two classes of problems. One is the variation of the opacity 
with time and position along and across the photosphere of our RSG; the second is the integration of the local line profiles along each line of sight 
and the addition of all the lines of sight over the stellar disk, what we can dub the geometry part of the radiative transfer problem. If we make 
the choice of doing full radiative transfer on a numerical simulation, both aspects of the problem, the geometry and the opacity, are inevitably 
together and we loose perspective on their relative importance. As it comes, we shall see that the geometry is mostly responsible of the main features 
observed in the atomic line profiles of Betelgeuse, playing opacity variations a second role. 

We therefore reduce the transfer problem to its bare bones. We assume that the opacity is constant along the line of sight and all over the stellar 
disk, in spite of the obvious presence of density and temperature variations. We assume a normalized continuum emitted below the region of 
formation of the line. We parameterize this region of formation with a geometrical distance $z$ which will vary from 0 through 1, $z=0$ being at 
the strict bottom end of the region of formation, and $z=1$ being at the strict top end. At each line of sight, the emergent spectrum will look like
\begin{equation}
   I(\nu)=e^{-\tau(\nu)}  
\end{equation}
where $\tau(\nu)$ is the total opacity along the line of sight, and we chose to parameterize wavelength in terms of velocity differences $v$ respect 
to the local reference frame of the star's center of mass  $v_*$ (where $v=0$\,\kms).The total opacity is computed as 
\begin{equation}
   \tau(\nu,\theta,\chi)=\int_{z=0}^{z=1} k \phi[\nu-v(z,\theta,\chi )\cos \theta]dz
   \label{opacityintegral}
\end{equation}
where $k$ is the constant absorption coefficient, and $v(z,\theta,\chi)$ is the radial velocity of the atoms at height $z$, angular distance to the
disk center  $\theta$ and position angle $\chi$. Finally we approximate the line profile $\phi$ by a simple gaussian of standard deviation $\Delta=6$\,\kms \citep{lopez_ariste_convective_2018}
\begin{equation}
   \phi(\nu)= e^{-\frac{\nu^2}{\Delta^2}}.
\end{equation}
All those local intensity profiles, integrated over $z$, for each position over the disk $(\theta,\chi)$ will be added together to form the 
disk-integrated line profile that we will compare to the observations.

Our approach is guided by the work of \cite{bertout_line_1987} (see also \cite{wagenblast_spectral_1983} and \cite{chandrasekhar_formation_1945}) to interpret the doubling profiles periodically observed in Mira 
stars. Those authors also strip radiative transfer to its bare fundamentals to demonstrate the somehow unexpected result of combining disk integration 
and velocity gradients along the line of sight. In their work they address Mira stars, whose atmosphere is supposed to be made of an expanding (or contracting)
shell of gas on top of the star. This shell is assumed to be homogeneous in velocity and geometry, what allows the integral over the disk to be 
intimately intricated with the integral along the line of sight. This helps in their solution to the problem. In our present case we shall also call 
for strong gradients of velocity along the line of sight, but these will only appear in coincidence with the plumes of hot, rising plasma, and will not be 
homogeneous nor in their distribution over the disk nor in their velocities. For this reason we shall decouple the integration over the line 
of sight from the integration over the disk. And in this aspect we diverge from the work of \cite{bertout_line_1987}. Despite that difference, we shall recover most 
of the features described by those authors.

\begin{figure*}
\includegraphics[width=0.3\textwidth]{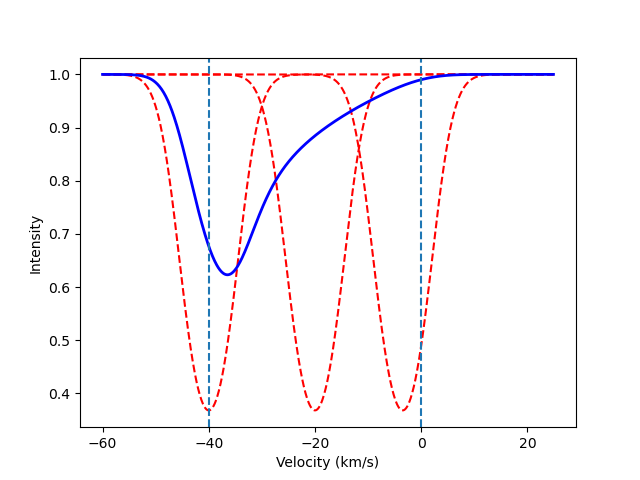}
\includegraphics[width=0.3\textwidth]{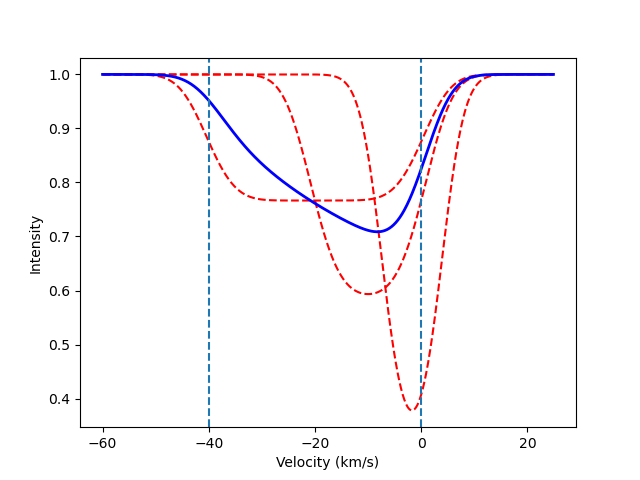}
\includegraphics[width=0.3\textwidth]{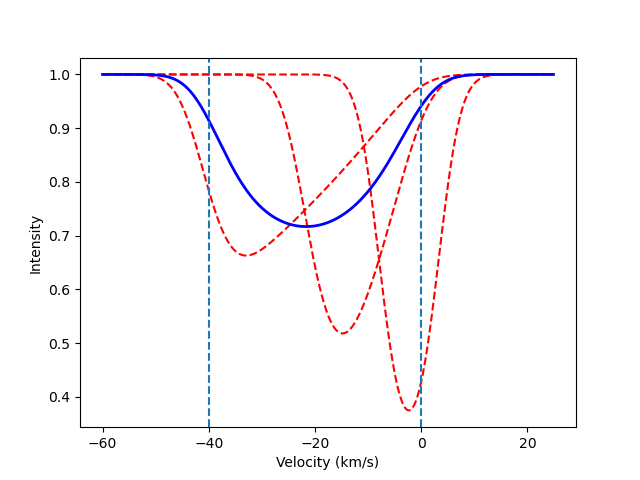}
\caption{Examples of  profiles generated in the presence of strong gradients along the line of sight, and the result of summing them up. 
On the left profiles made in absence of velocity gradients. The profection onto the line of sight shifts the position of the local profile. A simple limb-darkening 
law proportional to $\cos \mu$ makes asymmetric the summed profile. At center, a linear velocity gradient is used, while at right the
gradient  along the line of sight follows a square-root law, in all cases  with the same limb-darkening law as in the left. Dashed red lines are used for individual lines of sight from disk at 
disk center, $\theta =45^{\circ}$ and close to the limb. Continuous, blue lines show the disk-integrated profile.  The vertical dashed 
lines mark the radial outward velocity of the plasma $v_0=v_{\rm{max}}=-40$ \kms in the observer framework, and the zero velocity $v_*$.}
\label{example}
\end{figure*}

It is worth to solve analytically the integral in Eq. (\ref{opacityintegral}) for several gradients of the velocity with $z$.  In the first and straightforward case the 
velocity is $v_0$, constant with $z$ for all points in the disk, but projects itself onto the line of sight. We find that
\begin{equation}
   \tau(\nu,\theta)=k (z_1-z_0)\phi[\nu-v_0\cos \theta]
\end{equation}
As one adds profiles for different values of $\theta$ one recovers a flat-bottom or square integrated  profile. In the presence of  limb darkening this flat-bottom profile
becomes a highly 
asymmetric profile that can be seen at left, in Fig. \ref{example}. This is what one would expect from a constant 
and homogeneous rising velocity of the plasma. This is not what one sees, but yet one does not expect the velocity to be constant with $z$.

Hence, our second case assumes a linear dependence of the velocity with $z$ in the form $v(z)=v_0(1-z)$ which ensures that the velocity dimishes 
with height. The integral comes out to be
\begin{equation}
   \tau(\nu,\theta)=\frac{\Delta k}{v_0\cos\theta}\frac{\sqrt{\pi}}{2} \left[ {\rm erf}\left(\frac{\nu}{\Delta}\right) - {\rm erf}\left(\frac{\nu-v_0\cos\theta}{\Delta}\right) \right]
\end{equation}
where ${\rm erf}$ stands for the error function. Again we assume this law to be the same all over the disk, and the only angle dependece arises from 
the projetction onto the line of sight. Whenever $v_0$ is smaller than $\Delta$ this is just a slightly deformed gaussian that keeps adding 
up for all and every point over the disk. However if $v_0$ is larger than $\Delta$ we find that for large values of $\theta$, near the limb, quasi-gaussian 
profiles are produced, but near disk center, when $\cos \theta \approx 1$, flat-bottom  profiles appear. Adding up over all values of $\theta$ results 
in a disk-integrated  assymetric profile which can be seen in the center plot of Fig.\ref{example}

However a linear dependence of the velocity with $z$ is not what one naively expects from plasma balistically sent upwards during convection. One 
rather expects a square root dependence, $v(z)=v_0\sqrt{1-\beta z}$, where  $\beta$ modulates  the final velocity of the plasma. Such 
velocity can also be readily integrated and the total opacity per line of sight comes out to be
\begin{eqnarray}
   \tau(\nu,\theta)&=&\nu \frac{\Delta k \sqrt{\pi}}{\beta v_0^2\cos^2\theta} \left[ {\rm erf}\left(\frac{\nu-v_0\cos \theta\sqrt{1-\beta }}{\Delta}\right) \right.  \nonumber \\
   &&\left. -{\rm erf}\left(\frac{\nu-v_0\cos\theta}{\Delta}\right) \right]+\nonumber \\
   &&+\frac{k\Delta^2}{\beta v_0^2\cos^2\theta}\left(e^{-\frac{\left(\nu-v_0\cos\theta \sqrt{1-\beta }\right)^2}{\Delta^2}}-e^{-\frac{(\nu-v_0\cos\theta)^2}{\Delta^2}}\right)
\end{eqnarray}
One readily notices the linear dependence on $\nu$ on the first term of this solution, a fact already brought up by \cite{bertout_line_1987} 
in their geometry and which is maintained despite the divergences 
between our two approaches. The resulting profiles, if this term dominates over the second, show a characteristic saw-tooth or triangular shape that 
can also be seen in Fig.\ref{example}. The net result of summing up these and the usual gaussian profiles arising close 
to the limb is a broadened but quasi-symmetric profile.

When changing the gradients of the velocity along the line of sight, even with no modification of the maximum velocities involved (which is 
always 40\,\kms in the examples of Fig.\ref{example}) we witness evident changes in the shape of the profiles. The 
examples in Fig. \ref{example} show the path towards narrowing or broadening the disk-integrated line profile. 
Furthermore, and in advance of the results in Sect. 4, we also see how the bisector is also  modified in these tests, 
without requiring changes in neither the velocities or the absorption 
coefficient.  The three examples of Fig.\ref{example} show a typical C-shape bisector (no gradients case), an inverse C-shape bisector 
(linear gradients) or a quasi-vertical bisector (parabolic gradients).
In summary, the combination of strong gradients of different dependence with $z$, brightness 
inhomogeneities and disk integration 
appears to be what is needed to reproduce the observed intensity profiles in different RSG.

\section{Reproducing the observed intensity profiles.}
 
\begin{figure*}
   \includegraphics[width=\textwidth]{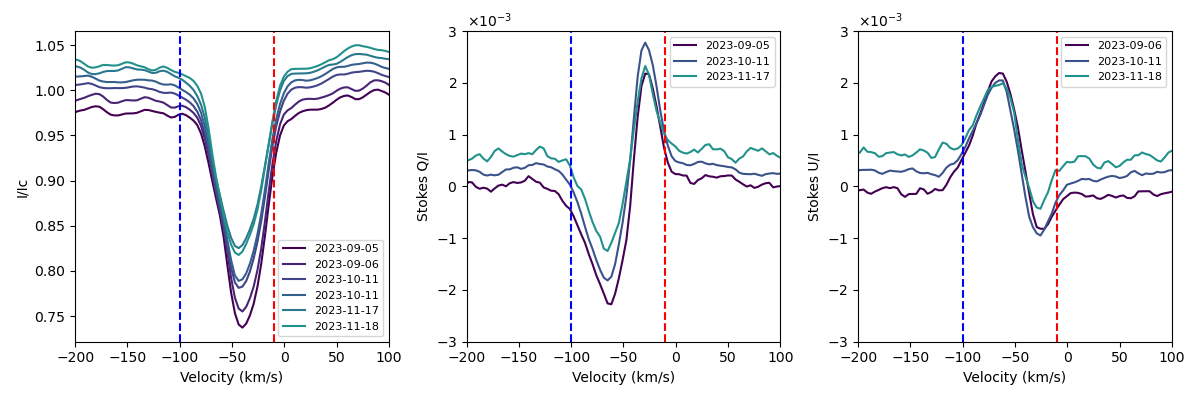}
   \caption{ Examples of profiles of RW Cep observed with NeoNarval in 2023 (dates indicated in the legend).  Intensity is on the left plot, Stokes Q and U are shown 
   in the center and right plots. The red and blue vertical dashed lines mark the chosen values of $v_*$ and $v_{max}$ respectively.}
   \label{observed}
   \end{figure*}

Figures \ref{obsprofiles} and \ref{observed} show selected examples of profiles of Betelgeuse and RW Cep respectively, which represent the two extremes of profile shapes 
observed in RSGs with Narval and its updated version, NeoNarval \cite[see][for a description of both instruments and the data reduction procedures]{lopez_ariste_three-dimensional_2022,donati_espadons_2006}.
What we are presenting in all our plots are cross-correlation profiles made by the addition of many different spectral lines  \citep{josselin_atmospheric_2007,donati_spectropolarimetric_1997}. Most 
often all those lines are added up irrespective of their line depression, formation height or any other 
attribute. This is what has been done in the case of the profiles shown in  Fig. \ref{observed}. In the case of Betelgeuse, the presented profiles date from 2013 and 2014. These data have been presented before by 
\cite{auriere_discovery_2016} who discuss also the line lists used in the cross-correlation. Though Betelgeuse is being observed periodically 
in recent years, after the great dimming event of the end of 2019 the levels of linear polarisation have drastically diminished. Since 
it is our purpose to illustrate the difference in wavelength span between the linear polarisation and the intensity profile, we have 
preferred the old data to better illustrate this discrepancy. However, note that for the Neo-Narval data in 2023 the broadening difference between Stokes Q,U and I remains unchanged.
 For our 
second star, RW Cep, we present NeoNarval data from 
2023, August 18th through November, 18th.  

We have seen that the presence of velocity gradients larger than the width of the line and of different dependences with $z$ opens the door to a 
variety of profile shapes, from purely gaussian to saw-tooth shapes, from broad to narrow ones. 
Which one is present in the observation 
of an RSG at a given date depends on the distribution of both gradients and brightness over the disk at the moment of the observation.
For the sake of simplicity we are going to fix gradients to be of the ballistic kind, with a square root dependence on $z$. But we are going 
to vary both the amplitude of the gradients (changing the parameter $\beta$) and the distribution of brightness and velocity amplitudes over 
the disk. The parameter $\beta$ will be set constant for all points in disk. For the brightness and velocity distributions, we will select the images inferred from linear polarisation of 
Betelgeuse to represent what we believe are acceptable  distributions over the disk, recalling that both magnitudes, brightness and velocity, are relied in the 
inversion algorithms fitting linear polarisation as if the light-emitting plasma was in the presence of convection \citep{lopez_ariste_convective_2018}. 

\begin{figure*}
   \includegraphics[width=0.8\textwidth]{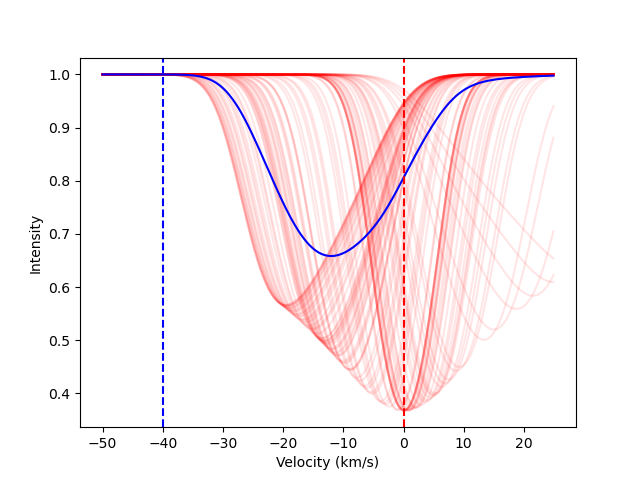} 
   \caption{ Example of local (red) and integrated (blue) profiles along the vertical radius of the image of Betelgeuse that fitted best 
   the observations of linear polarisation on 20th December 2013. All profiles are normalized to the continuum value for the sake of the visibility of the plot.}
   \label{figequateur}
   \end{figure*}

As a first illustration, using the image of the photosphere of Betelgeuse that best fits the observed linear polarisation signals observed on 
27 November 2013 \citep{auriere_discovery_2016} we present in Fig. \ref{figequateur} the local profiles along a radius across the stellar disk
together with the integrated profile summing up all those individual profiles. In that date, and along the chosen radius, there is no much contribution at the maximum 
velocities. There are however several points where the plasma is sinking and which can be seen redshifted respect to the $v_*$ velocity. In the figure, 
every local profile has been normalized, and these redshifted points are easily spotted. But they barely contribute to the integrated profile, for its true intensity is low. 

As described before, the profiles close to disk center contribute to the largest, blue shifted, velocities. But due to the velocity gradients they 
also contribute at lower velocities. Profiles close to the limb contribute only to the lower velocity, near the $v_*$ rest velocity of the star. 
Altogether wavelengths near this rest velocity get a larger contribution in the integration, and the net result is a narrow profile blue-shifted but with the rest velocity $v_*$ found in its red wing.

\begin{figure*}
   \includegraphics[width=0.8\textwidth]{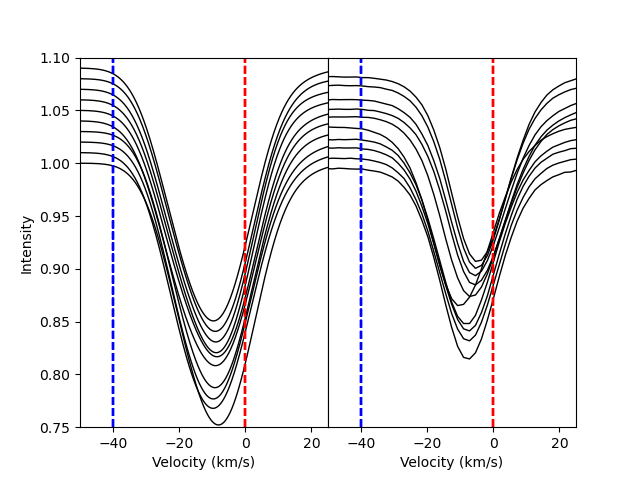}
   \caption{Left: Inferred images of Betelgeuse from 27 November 2013 through 3 March 2015 are used to integrate profiles assuming that at every point 
   the gradient is such that the velocity reduces to 0 at the top of the formation region of the line ($\beta=1$). Profiles are shifted in ordinates 
   for clarity. Right: The actual observed intensity profiles in those same dates.}
   \label{betelbeta1}
   \end{figure*}

We can now, extend the calculation to the whole disk and explore the impact of the velocity gradients, Fig. \ref{betelbeta1} shows the case 
of $\beta=1$, that is the velocity at each point drops from the value found in the inferred image to zero 
in the span of the formation region of the line. This is quite a large gradient near disk center, with maximum velocities reaching 40\,\kms but the gradients reduces to  zero towards the limb. The disk-integrated profile will be a combination of saw-tooth and gaussian 
profiles as in Fig.\ref{example}. But the distribution of brighness is not homogeneous, and some profiles will be given more weight 
than others. The result can be seen in Fig. \ref{betelbeta1} where  10 observations, from 27 November 2013 through 3 March 2015, presented by 
\cite{auriere_discovery_2016} are used as examples of brightness and velocity spatial distributions. The noticeable fact on this figure 
is that similar widths are seen on both the observed and computed profiles (of the order of 30km/s).

We have chosen the opacity value $k$ in Eq. (\ref{opacityintegral}) so that the disk-integrated profile matches the depth of the observed 
intensity profile. Otherwise all parameters are those used by the inversion code to fit the linear polarisation profiles: maximum velocity 
amplitude and width of the Gaussian profile. The comparison of these so-computed profiles 
and the observed ones is remarkable in terms of width and global shape. The computed profiles are narrow, in spite of the span of velocities used for the computation. This 
is due, as we keep repeating, to the combination of large velocity gradients and the brightness distribution over the disk. The zero velocity
falls in the red wing of the line rather than on line-center. This justifies \textit{a posteriori}  the ad-hoc choice made in the inversion 
codes by \cite{lopez_ariste_convective_2018}. Here, this is an emergent feature, there is 
nothing in the computation that forces the position of the minimum of the disk-integrated profile. The same can be said about the 
maximum velocity, that falls over the continuum well beyond the blue wing of the line, in spite of the presence of large velocities 
at some points over the disk. We conclude from this comparison, that our basic model captures the main features of the observed profiles 
of Betelgeuse. This is in spite of imposing a constant absorption coefficient over the atmosphere, and of imposing the same velocity 
gradient dependence for all points.  We shall look later into the bisector shape and the velocity spans to further push this comparison. 
At this point we can however dare to extract a far-edged conclusion. The model for linear polarisation is based upon the assumption that 
this linear polarization emerges from Rayleigh scattering of the continuum. The continuum polarised photons are subsequently absorbed 
by atoms higher in the atmosphere which re-emit them unpolarised. The local continuum polarisation is in this manner deleted in the spectral line, and it is 
this de-polarisation signal that we measure with our instruments. While the polarization amplitude, in that model, is mostly attached 
to the brightness of the continuum emitting plasma, the Doppler shift of the de-polarized photon is that of the absorbing and re-emitting atom. 
Linear polarisation profiles present a broad span  of velocities, but the present comparison of computed and observed intensity 
profiles appears to require a gradient of velocities down to zero over the region of formation of the line. We must conclude 
in consequence that de-polarised photons are coming from the very bottom of this formation region, while the photons making 
the spectral line come from the integration along the whole region.

\begin{figure*}
   \includegraphics[width=0.8\textwidth]{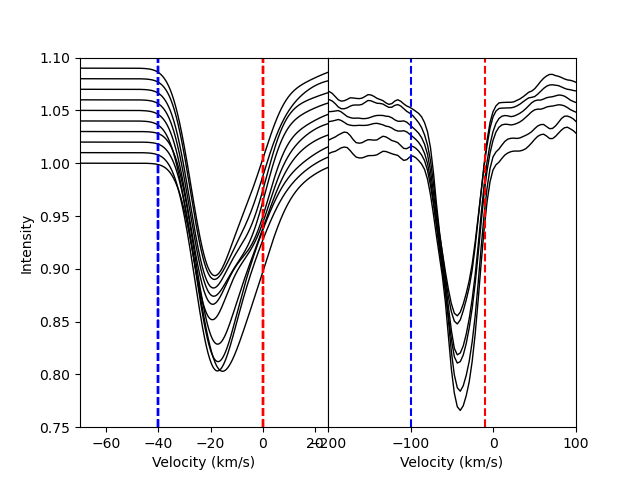}
   \caption{Left: Same as Fig. \ref{betelbeta1} but for $\beta=0.4$. The maximum velocity has been kept at 40\kms as in Betelgeuse. Right: Observed intensity profiles of RW Cep  as in Fig.3. The points of comparison 
   are the broadening of the profiles and the triangular shape illustrated by the step-rising blue wing of the profiles.}
   \label{RWbeta}
   \end{figure*}

RW Cep presents the other end of the variation seen in the intensity profiles of RSGs. We see,in the recent observations of this star, 
broadened profiles with a span comparable to the one observed in the linear polarisation profiles (Fig. \ref{example}). 
\cite{josselin_atmospheric_2007} also 
presented intensity profiles for RW Cep, though without linear polarisation, that can be compared to the ones presented here to appreciate 
all the variability of which this RSG is capable, unlike the stable profiles of Betelgeuse.
In Fig.\,\ref{RWbeta} we repeat 
the  calculations of disk-integrated profiles with the same brightness and velocity distributions used for Betelgeuse in Fig.\,\ref{betelbeta1}
(and therefore inferred  
from the fit of the observed linear polarisation of Betelgeuse), just changing the value of $\beta$ from $1$ into $0.4$. This value of $\beta$ means 
that rather than dropping to zero along the formation region of the spectral line, the velocity diminishes by just 25\%. With such 
small gradients we recover the broadened profiles that we can favourably compare to the observed profiles of RW Cep. Whenever 
the brightness distribution is such that hot large spots are present over the disk, one can even find split profiles. These are not 
seen in the present observations of RW Cep.

This second favourable comparison between the broadened profiles resulting from weak gradients in our simple model and the observations of RW Cep gives further support to it. In our view it confirms 
the hypothesis that the geometry of the radiative transfer and, in particular, the combination of the brightness distribution and 
the actual gradients of velocity present at the particular moment of the observation, are sufficient to qualitatively reproduce 
the observed intensity profiles. 

The scenario for the polarized line formation that emerges from these tests is the following. A continuum forms deep in the atmosphere of Betelgeuse (or any other RSG). This continuum is 
polarized by Rayleigh scattering and it is at this point that the information on brightness inhomogeneities is dump into the polarization amplitude. 
This does not have to be the continuum that finally emerges from the star: the optical depth in continuum wavelengths may not yet be zero.  It is depolarized by the line-forming atoms 
at the deepest point of its formation region. This has to be  so because we observe that linear polarization signals do not suffer from integration along the line of sight, 
and  they conserve  the signature of the largest velocities found at the deepest point of the line formation region. This is coherent with the hypothesis of a single 
scattering event in the creation of the polarization signal. It is at this point in the optical path, where the line stars forming depolarizing a continuum which is itself still forming,
that the wavelength position of the polarization signal is acquired and, from it, the position over the stellar disk to which this polarization will be assigned. Both the
 brightness of the image inferred from linear polarization 
and its position over the disk would come, in this scenario, from deep photospheric layers: the brightness from a deep continuum, well beyond the 
typical values of opacity of the emergent continuum, and the position over the disk from the very first scattering on a line-forming atom at the bottom 
of the formation region.   From this point on, there is no further change in the linear polarization profile, but the intensity profile keeps changing 
as it moves along the line formation region taking its definitive triangular or gaussian shape at the top of the formation layer, when atmospheric velocities 
have been greatly reduced.

\section{Line Bisectors}

We pursue this exploration of the shape of the emergent intensity profiles by looking into one further observed feature: the bisectors 
of the intensity line profiles.

\begin{figure}
   \includegraphics[width=0.5\textwidth]{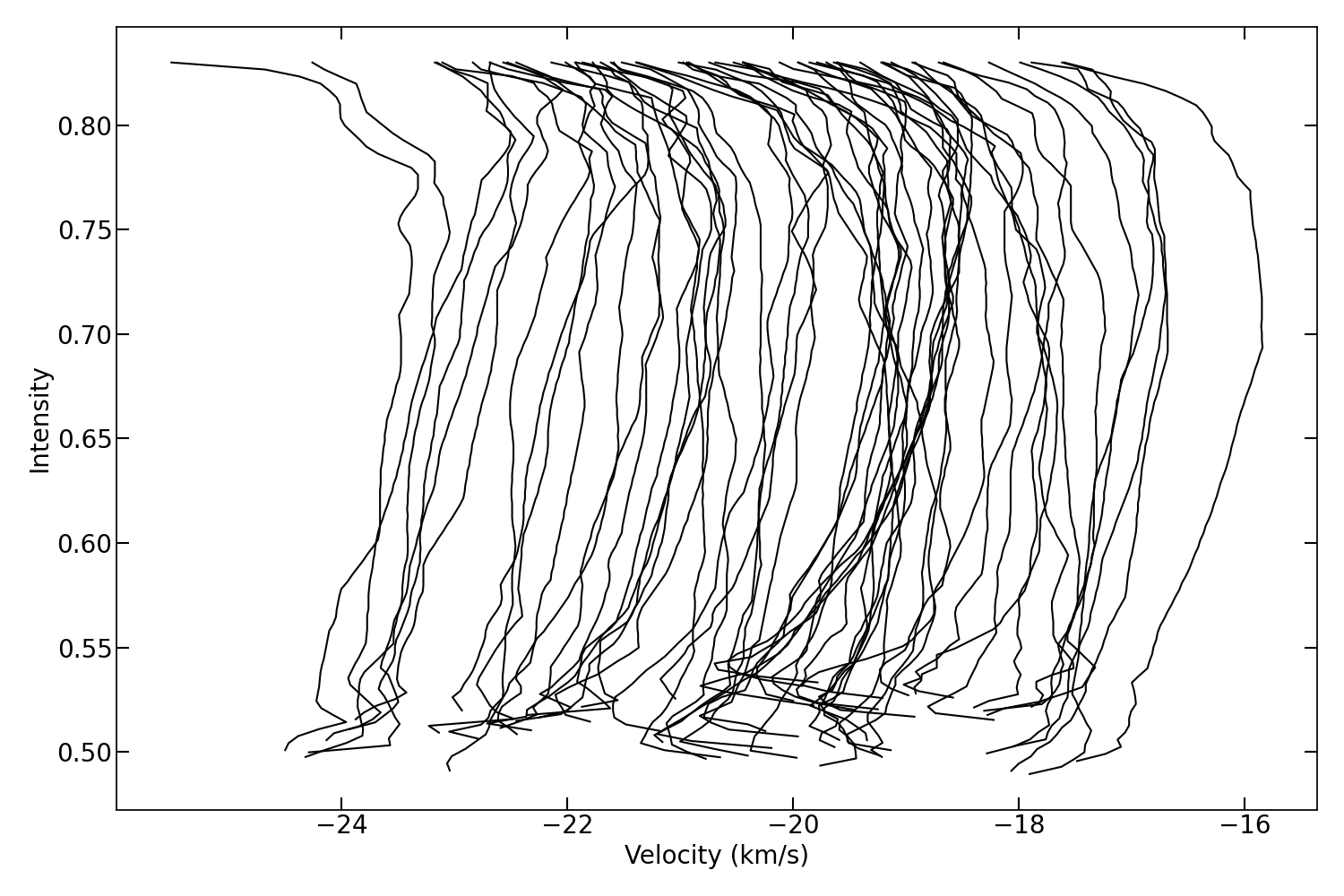}
   \caption{Measured bisectors in the cross-correlated intensity line profile of spectra of Betelgeuse observed from  November 2013 through April 2019 with Narval at the TBL. 
   The cross-correlation
   uses Mask C5 from \cite{kravchenko_tomography_2018} that selectes lines forming high in the photosphere of the star.}
   \label{couche0}
   \end{figure}

\begin{figure}
      \includegraphics[width=0.5\textwidth]{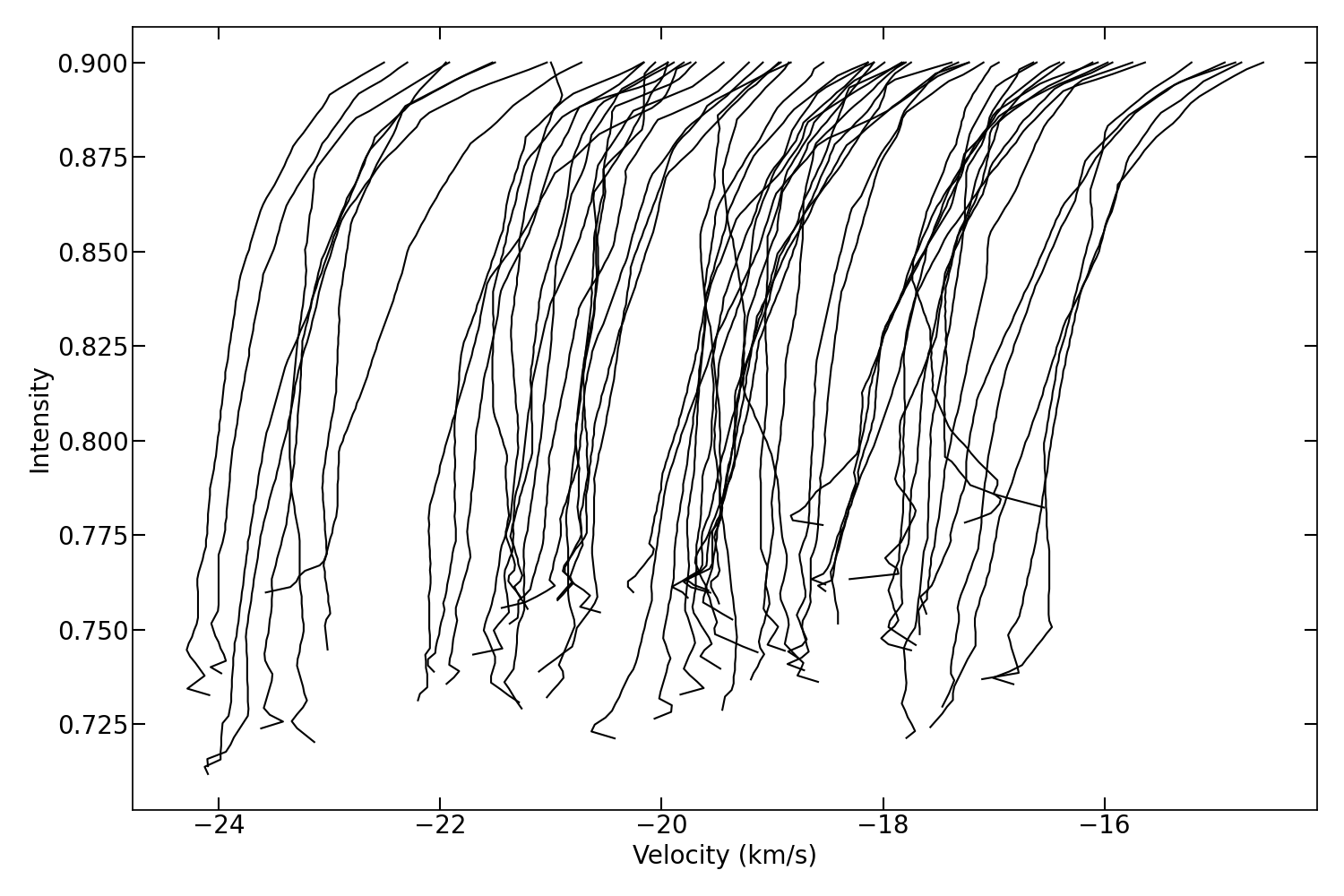}
      \caption{As in Fig. \ref{couche0}, but the mask used for cross-correlation selects atomic lines with central depressions in the range 0.7 through 
      0.8 of the continuum. This filter approximately selects lines forming deep in the photosphere of the star.}
      \label{couche6}
      \end{figure}

Figures \ref{couche0} and \ref{couche6} show the measured 
bisectors on the intensity profiles of Betelgeuse during the period November 2013 through April 2019. In Fig. \ref{couche0} the cross-correlated intensity profiles are 
made with a line mask  that selects lines formed in the upper part of the atmosphere. This mask was selected and created by 
 \cite{kravchenko_tomography_2018}. Strongly curved 
inverse C-shape bisectors are seen in the lines formed in this top photospheric layer, a feature traditionally interpreted as the 
presence of accelerating rising plasma.   In Fig. \ref{couche6} lines forming deeper in the photosphere have been selected, by the simpler 
procedure of selecting lines with central depressions in the range of 0.6 to 0.7 the continuum intensity. In these deepest layers, and 
for all the periods observed, the bisectors are straight with a tendency to bend towards the red in the top parts, in what looks as 
the more common C-shape bisectors.

\begin{figure*}
   \includegraphics[width=0.8\textwidth]{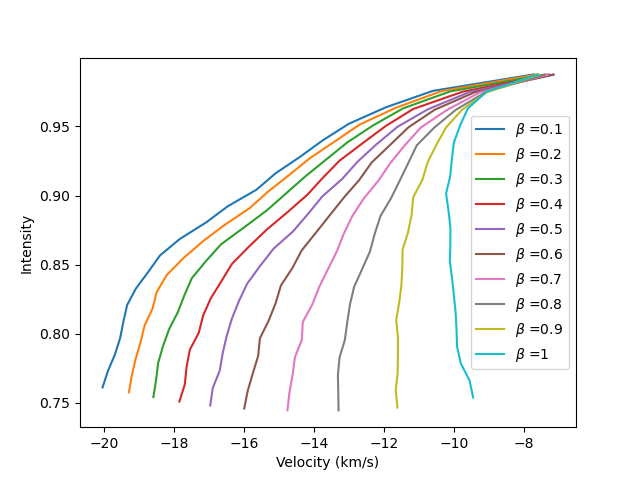}

   \caption{Bisectors computed using the image of Betelgeuse inferred to best fit the linear polarisation signal observed on 20 December 2013. Parabolic 
   gradients reducing the radial velocity are using with various values of $\beta$ as indicated. The maximum velocities at each point over the disk
   have been reduced to 70\%  the inferred value from the fit of linear polarisation signal to avoid the broadening of the profiles in the cases with 
   small $\beta$, thus keeping the computed profiles narrow, as the observed ones. }
   \label{bisector1}
   \end{figure*}

Reproducing not just the width of the profile, as done in the previous section, but also the bisector asymmetry is harder with such 
a simple radiative transfer model as the one  we  have chosen. Before, the presence of strong gradients was seen as a requirement to make 
compatible the broad linear polarisation profiles and the narrow intensity profiles of Betelgeuse, while reducing those gradients was the
simple change  required to also explain the broader RW Cep profiles. This main conclusion also applies when trying to reproduce bisectors.
 Fig. \ref{bisector1} shows 
the bisectors of disk-integrated profiles using the inferred image of Betelgeuse for the date of December, 20th 2013. The square root gradient law
is used as before in all cases. However, the value of $\beta$ is changed to simulate several situations: from almost no gradient ($\beta=0.1$) to a
 full decrease to zero  velocity along the line formation region ($\beta=1$). Affording very low gradients while maintaining the gaussian shape 
 of Betelgeuse profiles and not drifting towards 
RW Cep-like profiles requires reducing the maximum velocities to 70\% the original value, from 40 to 30 \kms. With this difference taken into account, 
the computed bisectors can be favourably compared to the observed ones of Fig.\ref{couche6}, particularly for values of $\beta \sim 0.5$ through $0.8$. 
Such lower values of $\beta$ may be justified by the smaller region covered by the formation region of these lines, compared to the full atmosphere 
encompassed when all the lines available are used to compute the cross-correlation profile. But one must 
not over-interpret the comparison of real data with such a simple radiative transfer problem. The important conclusion is that 
observed bisectors are still sufficiently well reproduced by this simple model requiring strong gradients and the actual plasma velocities 
and brightness inhomogeneities inferred from the fit of the linear polarisation profiles. 

\begin{figure*}
   \includegraphics[width=0.8\textwidth]{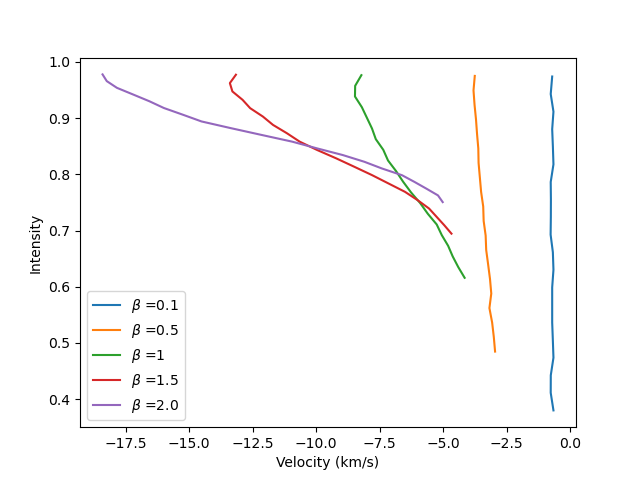}
   \caption{Computation of bisectors as in Fig. \ref{bisector1} but with a linearly accelerating gradient. Several cases of  spatial acceleration $\beta$ 
   are presented as indicated in the labels.}
   \label{bisector2}
   \end{figure*}

The lines formed in the top photospheric layers present in Fig.\ref{couche0} a pronouncedly inverse C-shape at all dates. These lines  dominate at particular dates 
the total  cross-correlation line profile. This global profile  may present at particular dates an inverse C-shape,
though often it  presents a normal C-shape \citep{gray_mass_2008}. But when isolated in separate layers, the bisectors of lines forming in either layer 
maintain their shape at all dates. This inverse 
C-shape  we observed in lines forming in  the top photospheric layers is traditionally interpreted as the presence of accelerating and rising plasma. And so 
appears to be the case in our simple model  as well. To reproduce these inverse-C bisectors we need to abandon the square root decrease in velocity and 
use a gradient that increases  the velocity, for example, linearly as $v(z)=v_0\beta z$. The resulting bisectors can be seen in  Fig. \ref{bisector2}. For constant velocities 
($\beta=0.1$) the bisector is almost straigth. As the gradient increeases and, particularly, for $\beta >1$ inverse C-shape bisectors 
appear. These do not reproduce completely the observed bisectors of Fig.\ref{couche0}. One is tempted to suggest that the lines 
  in the upper photosphere start forming while negative gradients are still present, but that mid way along their formation region the gradient 
  changes sign and actual plasma acceleration appears.

\section{Conclusion}

A bare-bones radiative transfer scheme appears to be able to describe the main observed features of intensity line profiles in Red Super Giants (RSG). 
The main feature addressed in this work is the, at present unexplained, narrowness of the observed intensity line profiles when compared to the 
broad profiles seen in linear polarisation.  
Stripping the radiative transfer problem from most of its ingredients and, particularly, of any variation of the absorption coefficients 
along the optical path, or spatially across the stellar disk, allows us to distille the main responsible behind this difference between the profiles.
 We find 
that the combination of strong velocity gradients along the line of sight together with the brightness inhomogeneities of the stellar disk suffice 
to broaden or narrow the line profiles at will.  This can be easily understood with the following simplified picture: Strong velocity gradients make
 the points around disk center to contribute to the integrated 
line profile with triangular profiles spanning the whole range of velocities but more intense towards redder wavelengths; points close to the limb 
contribute 
narrow gaussian profiles at those same red wavelengths. The result is a narrow near-gaussian profile. It suffices to reduce the gradients in velocity 
to change the triangular profiles arising close to disk center into bottom-flat square profiles that broaden the resulting profiles and can even 
split the intensity line. The linear polarization profile on the other hand arises from single scattering events at the bottom of the line formation 
region where the velocities are larger, and no radiative transfer takes places afterwards. The wavelength position of each linear polarisation signal 
is defined deep in the line formation region, and the signal amplitude reflects the continuum brightness seen by atoms deep in the photosphere, hence 
coming from even deeper atmospheric regions. This scenario appears to justify the validity of the brightness-velocity relationship imposed to the 
inversion algorithm that produces images of the photosphere of Betelgeuse from the linear polarisation profiles, a relationship only valid if convection 
still dominates plasma dynamics.

The use of this simple radiative transfer model together with the disk brightness distributions inferred from the fit of the observed 
linear polarization profiles in Betelgeuse produces intensity profiles which are remarkably analogous to the observed ones in terms of width and bisector shape.
 The very same model and 
brightness distributions can produce intensity profiles favourably comparable to the observed ones of RW Cep by just simply altering the gradient 
of the velocity. These two RSG appear to represent the extreme cases of the variability of the intensity profiles: Betelgeuse has shown narrow profiles 
for the last 11 years, while RW Cep shows broad intensity profiles. Both stars show similar broad polarization profiles.

The favourable comparison of intensity profiles, computed and observed, is comforted by the measure of the line bisectors. Once again, our simple radiative 
transfer model succeeds, by just tweaking the velocity gradients, in reproducing both the C-shape (decelerating velocity gradient) and inverse C-shape (accelerating velocity gradient)
 of the observed bisectors. The 
strongest inverse-C shape bisectors, observed in lines forming high in the atmosphere, appear to require a change of sign in the gradient. The decrease 
in velocity with height that justifies the C shape bisectors needs to be transformed into an acceleration to explain the inverse C shape. This may 
be (over) interpreted as these high lines being formed at regions where the stellar wind acceleration has already started.  

Summing up, the  narrow and broad intensity profiles of Betelgeuse and RW Cep  respectively  observed in the presence of broad linear polarisation 
profiles similar for both stars, do not require any change in the physics of these two red supergiants. Taking into account velocity gradients along the line of sight 
and integrating over an inhomogeneous disk suffices. Strong gradients explain the narrow intensity profiles of Betelgeuse, smaller gradients explain 
the broad profiles of RW Cep. The hypothesis of single scattering and correlation of velocity and brightness, used for the interpretation of the 
linear polarisation, are comforted by our results.

If our conclusions are correct, RW Cep is at present not capable of diminishing the large vertical velocities of its photospheric plasma. These large 
velocities may help larger amounts of plasma escape the gravity and we may expect RW Cep to spell larger clouds of circumstellar matter. Betelgeuse on the 
contrary is mostly able to slow down plasma and its stellar wind must be dwindling. How two RSG, mostly identical in their fundamental parameters, can 
behave so differently is an open question. The apparent existence of cyclical Decin stages \citep{decin_probing_2006,lopez_ariste_height_2023} supports the idea that these larger or smaller velocity 
gradients appear repeatedly in the life of a Red SuperGiant.

\begin{acknowledgements}
This work was supported by the "Programme National de Physique Stellaire" (PNPS) of CNRS/INSU co-funded by CEA and CNES.
We acknowledge support from the French National Research Agency (ANR)
funded project PEPPER (ANR-20-CE31-0002)
\end{acknowledgements}

\bibliographystyle{/Users/art2/TeX/aanda/bibtex/aa}

\bibliography{art74}

\end{document}